\documentstyle[preprint,aps]{revtex}
\begin{document}
\draft
\title{Resonance Magnetoresistance in Coupled Quantum Wells}
\author{Y. Berk $^1$, A. Kamenev$^2$, A. Palevski$^1$,
L. N. Pfeiffer$^3$, and K. W. West$^3$ }

\address{$^1$School of Physics and Astronomy, Raymond and Beverly Sackler
Faculty of
Exact Sciences,\\
Tel Aviv University, Tel Aviv 69978, Israel\\
$^2$Department of Condensed Matter, The Weizmann Institute of Science,
Rehovot 76100, Israel.\\
$^3$ AT\&T Bell Laboratories, Murray Hill, New Jersey 07974}

\date{\today}
\maketitle

\begin{abstract}
The in-plane magnetic field suppresses the quantum coupling between
electrons in a double quantum well structure. The microscopical
theory of this effect is developed and confirmed experimentally.
We have shown that the decrease of the "resistance resonance" peak is
sensitive to the mutual orientation of the  current
and the in-plane magnetic field.
The characteristic field required for the suppression of the resonance
depends on the elastic small angle and electron--electron scattering rates.
The  study of the characteristic field
allows to verify the temperature and Fermi energy dependence of  the
electron--electron scattering rate, providing a new experimental tool for
its measurement.

\end{abstract}


\narrowtext

A new physical phenomena, called resistance resonance (RR), in a double quantum
well (QW) structure was recently predicted and observed experimentally
\cite{Palevski}.
The key point of this effect is the following. Let us consider two {\em
tunneling
coupled} QWs. The  quantity of interest is the {\em lateral}
resistance of the structure (all the electrodes
are attached to the both QWs !). If the tunneling between QWs is by some
reason suppressed, each electron is localized in one of the wells.
The resulting
lateral resistance is those one of two  conductors connected in parallel,
$R_{off} \sim (\tau^{tr}_1 + \tau^{tr}_2)^{-1}$, where $\tau^{tr}_i$
is the transport  mean free time in the $i^{\underline{th}}$ well. In the
presence of tunneling, the eigenfunctions form
symmetric and antisymmetric subbands, leading to delocalization of electrons
between two wells.  The corresponding scattering
rate in each of these subbands is
$(\tau^{tr})^{-1} = (2\tau_1^{tr})^{-1} + (2\tau_2^{tr})^{-1}$ and
the resistance is given by $R_{res} \sim (2\tau^{tr})^{-1}$. One notices
that if mobilities of the two QWs are different
($\tau_1^{tr}\neq\tau_2^{tr}$), than $R_{res} > R_{off}$.
The reason is very simple: in the first case (no coupling) the clean well
shunts the dirty one, making the resistance small. No such shunting occurs
for the coupled wells.

The experimental realization of this idea \cite{Palevski,Pal,Sakaki}
 was based on the displacement of the QW's energy levels by the gate voltage.
The typical graph (lateral resistance vs. gate voltage) is plotted on the
insert to Fig. 2. The resonance occurs in the point, where the energy levels
of the two wells coincide facilitating the tunneling. In the present latter we
propose a different realization of the RR. Namely,  the
RR  is observed as a function
of the {\em in--plane} magnetic field
(instead of the gate voltage). The maximal resistance corresponds to the
zero magnetic field. The suppression of tunneling between two QWs
in the parallel magnetic field was already demonstrated (in a different
context) in Refs.\ \cite{Eisen} (perpendicular transport) and \cite{boeb}
(Shubnikov--de Haas oscillations). We employ the
nice intuitive picture, developed in Refs.\ \cite{Eisen,boeb}, to illustrate
the results of our calculations.
Below we present a microscopical description of the lateral
magnetoresistance of the coupled QWs, which is verified by the experimental
data. The main messages, following from our studies, are:
\newline (i) The in-plane magnetic field destroys the coupling between
QWs, leading to the RR;  the lateral resistance is essentially
{\em anisotropic}, means that  the shape of the RR depends on the angle
between the current and magnetic field.
\newline (ii) The {\em width} (i.e. the characteristic magnetic field, $H_c$)
of the RR is sensitive to the single electron scattering time, providing
a new method of measuring the small angle
scattering time on the remote impurities.
\newline (iii) The dependence of $H_c$ on temperature and on Fermi energy
suggests
that the electron--electron scattering rate (intralayer and interlayer) may
be tested as well.

To develop a microscopical model of transport in two
QWs we employ the basis of eigenstates of uncoupled wells. In this basis
the Hamiltonian of the system is a $2\times 2$ matrix, the off--diagonal
elements of which represent the tunneling coupling between QWs
\begin{equation}
\hat H_{\bf k,p}= \delta_{\bf kp}  \left( \begin{array}{lr}
({\bf p}-e/c{\bf A}_1)^2/(2m^*)   & \Delta/2 \\
\Delta/2  & \hskip -2cm      ({\bf p}-e/c{\bf A}_2)^2/(2m^*)
\end{array} \right)  + \left( \begin{array}{cc}
U_1({\bf p-k})   & 0  \\
0                &  U_2({\bf p-k})
\end{array} \right) \, .
                                                                \label{ham1}
\end{equation}
We treat here only the case of coinciding quantized (in the $z$ direction)
energy levels of the two wells.
In Eq.\ (\ref{ham1}) -- ${\bf k,p}$ are 2D  momentum of the electrons,
$\Delta$ is the
tunneling gap (we assume tunneling to be  momentum conserved)
and ${\bf A}_i $ is a  vector potential of an external field in the $i^{th}$
QW.
The second matrix on the r.h.s. of Eq.\ (\ref{ham1}) represents an elastic
impurity scattering inside each QW. We  assume that (1) random potentials
$U_i({\bf p-k})$ have a finite correlation length in a plane of 2D gas, (2)
there are no correlations between scatterers in different wells. In this
case the disorder potential (in each well) may be described  \cite{Abrikosov64}
by the single particle (small angle) mean free time, $\tau_i$, and the two
particle (transport) mean free time, $\tau_i^{tr}\geq\tau_i$.

In a uniform
magnetic field, $H$, parallel to the plane of the QW's (say directed along
the $y$ direction) the corresponding vector potentials are
${\bf A}_{i} = (H z_i, 0)$,
where $z_i$ are the $z$ coordinates of the effective centers of the QWs.
The Fermi surfaces of two QWs have a form of two  circles displaced along the
$x$ direction  (see Fig.\ 1)
on the relative distance $k_H=e/c\, H b$, where $b=z_1-z_2$ is
the distance between the centers of the wavefunctions in two wells
\cite{Eisen,boeb}. Only the electrons, which occupy the states in the
vicinity of the (quasi)crossing points $A$ and $B$ (see Fig.\ 1), have the same
energy ($\epsilon_F$) and momentum in both wells and hence participate in the
tunneling. Several important conclusions follow immediately:
\newline (i) The  resistance approaches its off-resonance value
(decreases) as the magnetic field is increased.
\newline (ii) The characteristic scale of the magnetic field may be estimated
as $v_F k_H\approx\max\{\Delta, \hbar/\tau\}$ (see below).
\newline (iii) The lateral resistance of the system in the in--plane
magnetic field is {\em anisotropic}. Indeed, the transport in the
$x$ direction is
dominated by  states with large $k_x$, which  are practically
decoupled (cf. Fig. 1 ). As a result the perpendicular (to the direction of the
field) resistance is close to the off--resonance value. Contrary,
the transport in the direction of the field is mostly determined by the
states situated near the points $A$ and $B$ of Fig.\ 1.
These states are delocolized, making the parallel resistance closer to
the resonance value.
In other words, the suppression of the RR occurs in a different way depending
on the angle between the magnetic field, ${\bf H}$, and the current,
${\bf j}$, used to probe the resistance.

The detailed diagrammatic calculation, based on the Kubo formula \cite{Berk94}
leads to the following dependence of the resonance resistance  on the
in--plane magnetic field
\begin{equation}
R^{-1}(H)-R^{-1}_{off}=(R^{-1}(0)-R^{-1}_{off})f(H/H_c),
                                                               \label{res}
\end{equation}
where
\begin{equation}
f(x)=\frac{2(\sqrt{1+x^2}\, -1)}{x^2}
\left\{ \begin{array}{ll}
1;                          &  {\bf H}\,  ||\, {\bf j}, \\
(1+x^2)^{-1/2}; \hskip 1cm  &  {\bf H} \perp {\bf j}
\end{array} \right.
                                                               \label{f}
\end{equation}
and the characteristic field is given by
\begin{equation}
H_c=\frac{\hbar c}{e}\frac{1}{v_F\tau b}
\sqrt{1+\left( \frac{\Delta}{\hbar} \right)^2
\frac{\tau_1^{tr}+\tau_2^{tr}}{2}\tau},
                                                               \label{hc}
\end{equation}
finally $2\tau^{-1}\equiv \tau_1^{-1}+\tau_2^{-1}$.
The above relations are valid until
$H\approx H_F$, where $H_F\equiv 2\pi\hbar c/(e\, \lambda_F b)$;
$\lambda_F $ is a Fermi wavelength. Note also that
$H_c\ll H_F$, as $\epsilon_F\tau/\hbar\gg 1$.
In agreement with our expectation, the
RR is suppressed faster in the perpendicular configuration,
$R^{-1}(H)-R^{-1}_{off}\propto H^{-2}$ , whereas in the parallel
configuration  $R^{-1}(H)-R^{-1}_{off}\propto H^{-1}$, for $H_c\ll H< H_F$.

The double QW structure was grown on N$^+$ GaAs substrate
by molecular-beam epitaxy and
consists of two GaAs wells 139 $\AA$ width separated by a 40 $\AA$
Al$_{0.3}$Ga$_{0.7}$As barrier.
The tunneling gap for this structure is estimated as $\Delta=0.55$meV.
The electrons were provided by remote delta-doped donor layers set back
by 250 $\AA$ and 450 $\AA$ spacer layers from the top and the bottom
well correspondingly.
In order to obtain the difference in the mobilities, an enhanced amount of
impurities  was introduced at the upper edge of the top well (Si, 10$^{10}$
cm$^{-2}$). The schematic cross-section of the device may be found in Ref.\
\cite{Palevski}.
Measurements were done on
10$\mu$m-wide and 200 $\mu$m-long channels with Au/Ge/Ni Ohmic
contacts.  Top and  bottom gates
were patterned using the standard photolithography fabrication
method.
The top Schottky gate covered 150 $\mu$m of the channel.
The data were taken using a lock-in four terminal
techniques at $f$= 5.5 Hz. The voltage probes connected to the
gated segment of the channel  were separated by 100 $\mu$m.
The complementary measurements of the resistance and Hall coefficient
leads to  the following parameters of the
structure (as grown, i.e., $V_g = V_{bg}=0$ and $T=4.2K$):
$\mu _1 = 47,000 cm^2/Vsec$, $\mu_2 = 390,000 cm^2/Vsec$,
$n_1 = 4.7 \times 10^{11} cm^{-2}$, $n_2 = 2.5 \times 10^{11} cm^{-2}$.
The values of these parameters for each temperature and gate voltage
were determined independently and used for the fits (see below).

The variation of the  top gate voltage, $V_g$ (for a fixed bottom gate
voltage, $V_{bg}$), allows  to sweep the potential profile
of the QWs through the resonant configuration.
The  resistance vs. top gate voltage   ($V_{bg} = 1.5 V$ ; $T=4.2 K$ )
is plotted in the inset to Fig. 2.
The resistance resonance is clearly observed at $V_g\approx -0.2 V$.
The value of the resistance in resonance is ${R}_{res}= 65 \Omega$, whereas
the off--resonance value is estimated as ${R}_{off}= 32 \Omega$
(cf. Fig.\ 2, inset). Next we fix the gate voltges, corresponding to the exact
resonance position, and measure the resistance as a function of the in--plane
magnetic field.  Fig.\ 2 shows the behavior of the RR for the two orientations
of the magnetic field with respect to the direction of the  current
$({\bf H}\,  ||\, {\bf j}$ and ${\bf H} \perp {\bf j})$.
The experimental data clearly demonstate the  suppression of the RR by the
magnetic field, as well as the expected anisotropy.
In the perpendicular orientation the
resistance decreases faster than in the parallel one.
The theoretical curves, using Eqs.\ (\ref{res}), (\ref{f}) with
${H}_c= 0.44 T$ (this is the only fitting parameter !) are shown on the same
plot. For the perpendicular configuration, we obtained a perfect fit for the
magnetic fields up to $H\approx 3 T$ (note that $H_F\approx 4.2 T$ for
$\epsilon_F=15$ meV).

The situation is markedly different for the parallel configuration.
The fit to the data  is obtained only in the
narrow range of fields up to  ${H}_c$, at  high magnetic fields the
resistance does not approach the value ${R}_{off}= 32 \Omega$.
Moreover, a positive magnetoresistance contribution is well resolved.
Large positive magnetoresistance ($\propto H^2$) in the
parallel configuration was also observed for the one QW (the second well was
totally depleted by a large  negative voltage on the top gate).
We tend to attribute this positive contribution (hence the poor fit)
to some normal to the plane of QWs component of
the magnetic field, which is due to non perfect flatness of our structure in
one direction. In the following, we thus restrict ourselves mostly on the
perpendicular (${\bf H} \perp {\bf j}$) orientation.

We employ now Eq.\ (\ref{hc}) and the extracted value of  the characteristic
field, ${H}_c= 0.44 T$, to establish the small angle scattering time, $\tau$
(note that all other parameters entering Eq.\ (\ref{hc}) are known, see
above). As a result one has $\hbar/\tau$=1.7 meV  at T=4.2 K, which
implies the ratio between the transport and the small angle scattering times
to be equal to $\approx 3.2$.  Measurements of this ratio for different
values of the Fermi energy (see below) result in a slow decrease from 3.2 at
$\epsilon_F=15$ meV to 2.5 at $\epsilon_F=7$ meV.
These data are in a good
agreement with the one  measured, using Shubnikov--de Haas
oscillations, in the 2D gas with the similar mobility \cite{Cole}.
It becomes evident now why one should complicate the theory to account  the
long range nature of scatterers. The simpler theory with short range
scatterers only ($\tau_i^{tr}=\tau_i$) fails to explain quantitatively the
observed width of the RR. We
conclude thus that the suppression of the RR in the magnetic field gives rise
to a new and relatively simple way of measuring small angle scattering time.

We repeat then the same procedure (using $H_c$ as the only fitting parameter
and then extracting  $\tau$) for the perpendicular orientation at
different temperatures in the range between $4.2 \div 40 K$.
The experimental data and a set of theoretical curves is presented in Fig. 3.
The width of
the curves increases with temperature indicating the increase of  ${H}_c$.
The same type of data were obtained for different set of voltages applied to
the top and the bottom gates corresponding to the resonant conditions at
different Fermi energies. The values of the Fermi energy in the range
$7\div 15$ meV were investigated. These data were also analyzed in the same
fashion and the values of  $\tau(T,\epsilon_F)$ were determined (we stress
again that all relevant parameters, besides $\tau$,
were established independently for each value of $T$ and $\epsilon_F$).

On the Fig.\ 4 we plot in a logarithmical scale
$\epsilon_F(\hbar/\tau(T)-\hbar/\tau(0))$ versus
temperature for three different values of the Fermi energy. At small enough
temperature ($k_B T\ll\epsilon_F$) all experimental points collapse to the same
line. The slope of this line implies the quadratic temperature
dependence of the displayed quantity. This way  the following
relation is established
\begin{equation}
\frac{\hbar}{\tau(T)}-\frac{\hbar}{\tau(0)}
\propto \frac{(k_B T)^2}{\epsilon_F}
                                                           \label{int}
\end{equation}
Eq.\ (\ref{int}) suggests that the single particle scattering rate,
$\tau^{-1}(T)$, consists of the two parts: small angle scattering rate on the
remote impurities, $\tau^{-1}(0)$,  and the electron--electron (e -- e)
scattering rate, $\tau^{-1}_{ee}$. This is in contrast to the transport (two
particle) scattering rate, $1/\tau^{tr}$, which is practically not affected
(in the clean limit, see below) by the e -- e scattering (due to momentum
conserved nature of the latest). To verify this idea quantitatively we use
the result \cite{Quinn76} for the e -- e scattering rate in a {\em clean}
(the criterion is $\hbar/\tau(0)\ll k_B T\ll \epsilon_F$, which is fulfiled
in our case) 2D gas
\begin{equation}
\frac{\hbar}{\tau_{ee}}=(1+\xi)\frac{1}{\pi}
\frac{(k_B T)^2}{\epsilon_F}
(1+\ln 2+\ln\frac{\lambda_F}{\lambda_{TF}}-\ln\frac{k_BT}{\epsilon_{F}}),
                                                           \label{ee}
\end{equation}
where $\lambda_{TF}=276\AA$ is the Thomas--Fermi screening length in the
GaAs. We have introduced in Eq.\ (\ref{ee}) an additional factor $(1+\xi)$,
which intends to simulate intralayer and interlayer contributions to the e --
e scattering. In the original theory \cite{Quinn76} only one 2D gas was
considered thus $\xi\equiv 0$. In the case of two close QWs one expects that
$0<\xi<1$, depending on the ratio between screening length, $\lambda_{TF}$,
and the mean distance between the wells, $b$. This is indeed the case: the
best fit  to our data (the solid line in Fig.\ 4) is achieved by
Eq.\ (\ref{ee})  with $\xi=0.5$. We conclude thus that in our structure the
interlayer e -- e scattering rate is $0.5$ of the corresponding intralayer
value. This seems to be reasonable since the distance between the wells is
of the order of the screening length. To make  more
quantitative statements  --  theory of e -- e interactions in two
tunneling coupled QWs would be desirable.
Eq.\ (\ref{ee})  is valid only in the limit ${k}_B{T}\ll\epsilon_F$.
Therefore the  deviations of the
experimental points from the theory at high
temperatures  (especially for the smallest  $\epsilon_F$) are not surprising.
Our results may be considered as an other confirmation of the theory
\cite{Quinn76} in the range of relatively large temperatures. In a small
temperature regime the theory \cite{Quinn76} was excellently confirmed in the
interference experiment \cite{Yacoby94}.

The central point to all our discussions is the fact, that the transport
quantity (the resistance) of the structure appears to be sensitive to the
single particle scattering time (and not only to the transport one !). This
enables us to determine the small angle scattering time, $\tau(0)$
(from the low temperature measurements, $k_B T<\hbar/\tau(0)$) as well as the
e -- e scattering time, $\tau_{ee}$ (from the measurements at
$k_B T>\hbar/\tau(0)$). The further comparison with the theory leads to the
reasonable estimation of the ratio between intrawell and interwell e -- e
scattering rates. We believe thus that the RR in two coupled QWs with
different mobilities  provides  a powerful and relatively simple method of
measuring the small angle and e -- e scattering rates.
Some unresolved questions raised in the present latter
require further theoretical and experimental investigation of this
phenomena.

We have benefited from the useful discussions with A. Aronov,
O. Entin, V. Fleurov, Y. Gefen, D. Khmelnitskii, Y. Levinson and A. Yacoby.
The experimental research was supported by Israel Academy of Sciences
and Humanities.
A.K. was supported by the
German--Israel Foundation (GIF) and the U.S.--Israel Binational
Science Foundation (BSF).

\figure{Fig. 1.
Fermi surfaces of two QWs in the in--plane magnetic field. }

\figure{Fig. 2. The resonance resistance vs.  magnetic
field at 4.2 K : circles and asterisks  -- experimental
data for the perpendicular and parallel orientations correspondingly; solid
and dashed lines --  theoretical curves.
The inset -- lineshape of the RR vs. top gate voltage, $V_g$ ($V_{bg}=1.5$ V).
}

\figure{Fig. 3. The  resistance vs. magnetic field in the perpendicular
configuration at different temperatures: circles -- experiment; solid lines
-- theory
(${V}_{gb} = 1.5$ V; ${V}_{G} = -0.2$ V, corresponding to
$\epsilon_F = 15 meV$). }

\figure{Fig. 4. Normalized scattering rate:
$\frac{\epsilon_F}{11 meV}(\frac{\hbar}{\tau(T)}-\frac{\hbar}{\tau(T)})$
vs. temperature for different Fermi energies. The solid line:
$\frac{\epsilon_F}{11 meV}\frac{\hbar}{\tau_{ee}}$, where $\hbar/\tau_{ee}$ is
given by Eq.\ (\ref{ee}) with $\xi=0.5$. }


\begin{references}

\bibitem{Palevski} A. Palevski, F. Beltram, F. Capasso, L. N. Pfeiffer
and K. W. West, Phys. Rev. Lett. {\bf 65}, 1929 (1990).

\bibitem{Pal} A. Palevski, S. Luryi, P. L. Gammel, F. Capasso, L. N. Pfeiffer
and K. W. West, Superlattices Microstruct.{\bf 11}, 269 (1992).

\bibitem{Sakaki} Y. Ohno, M. Tsuchia and H. Sakaki, Appl. Phys. Lett.{\bf 62},
1952 (1993).

\bibitem{Eisen} J. P. Eisenstein, T. J. Gramila, L. N. Pfeiffer, K. W. West,
Phys. Rev.{\bf B44},6511 (1991)

\bibitem{boeb} G. S. Boebinger, A. Passner, L. N. Pfeiffer, K. W. West,
Phys. Rev. B {\bf 43}, 12673 (1991).

\bibitem{Abrikosov64} A. A. Abrikosov, L. P. Gorkov, and I. E. Dzyloshinski,
{\it Methods of Quantum Field Theory in Statistical Physics}, Prentice-Hall,
Inc.,
1963.

\bibitem{Berk94}Y. Berk {\em et al}, to be published.

\bibitem{Cole} P. T. Coleridge, Phys. Rev. B {\bf 44}, 3793 (1991).

\bibitem{Quinn76} G. F. Giuliani, and J. J. Quinn, Phys. Rev.{\bf B 26}, 4421
(1982). Note the missed factor of two in the final result. For the discussion
of this point see Ref.\ \cite{Yacoby94}.


\bibitem{Yacoby94}A. Yacoby, M. Heiblum, H. Shtrikman, V. Umansky, and D.
Mahalu, Semicon. Sci. Technol. {\bf 9}, 907 (1994).

\end{references}
\end{document}